\documentclass[aps,prl,reprint,superscriptaddress]{revtex4-2}

\usepackage{graphicx}%
\usepackage{amsmath,amssymb,amsfonts}%
\usepackage{ragged2e}
\usepackage[title]{appendix}%
\usepackage{xcolor}%
\usepackage[justification=justified,singlelinecheck=false]{caption}
\begin{document}

\title{Thermodynamic Entropy as Information\\
\normalsize A compression-based demonstration of the Shannon–Boltzmann equivalence in condensed matter}

%%=============================================================%%
%% GivenName	-> \fnm{Joergen W.}
%% Particle	-> \spfx{van der} -> surname prefix
%% FamilyName	-> \sur{Ploeg}
%% Suffix	-> \sfx{IV}
%% \author*[1,2]{\fnm{Joergen W.} \spfx{van der} \sur{Ploeg} 
%%  \sfx{IV}}\email{iauthor@gmail.com}
%%=============================================================%%

%\author*[1]{\fnm{Dallin} \sur{Fisher}}\email{djfishe7@asu.edu}
\author{Dallin Fisher}\email{djfishe7@asu.edu}

%\author*[1]{\fnm{Qi-Jun} \sur{Hong}}\email{qhong@alumni.caltech.edu}
\author{Qi-Jun Hong}\email{qhong@alumni.caltech.edu}
\affiliation{Materials Science and Engineering, Arizona State University, Tempe, Arizona 85285, USA}

%\affil*[1]{\orgdiv{Material Science and Engineering}, \orgname{Arizona State University}, \orgaddress{\street{Street}, \city{City}, \postcode{100190}, \state{Az}, \country{USA}}}

%%==================================%%
%% Sample for unstructured abstract %%
%%==================================%%

%\abstract{
\begin{abstract}
We demonstrate that Shannon’s information entropy and the thermodynamic entropy of Boltzmann and Gibbs are quantitatively equivalent for real condensed-matter systems.  By interpreting atomic configurations as information sources, we compute entropy directly from the compressibility of molecular-dynamics trajectories, without physical partitioning or empirical modeling.  A custom lossy-compression algorithm measures the minimum number of bits required to describe a microstate at finite precision, and this bit count maps exactly to thermodynamic entropy through the Shannon–Boltzmann relation.  The method reproduces benchmark entropies for metals, semiconductors, oxides, and refractory ceramics in both solid and liquid phases, establishing information as the fundamental quantity underlying thermodynamic disorder.  This equivalence unifies information theory and statistical mechanics, providing a general and computationally efficient framework for determining entropies and free energies directly from atomic data.
%Entropy links the microscopic and macroscopic worlds, yet its direct evaluation in materials remains computationally demanding and conceptually opaque. Conventional approaches decompose the total entropy into vibrational, configurational, electronic, and magnetic terms, each requiring separate physical models and approximations. Here we take a different route: we implement, literally and computationally, the equivalence between Shannon and thermodynamic entropy. Using only the information content of atomic configurations from first-principles molecular-dynamics simulations, we compute entropy without invoking any explicit physical models. A custom lossy compression algorithm measures the minimum number of bits needed to describe a microstate within a defined precision, which maps directly to the thermodynamic entropy using the Shannon–Boltzmann relation. Applied to simple metals and semiconductors, this information-theoretic framework reproduces benchmark SLUSCHI/mds results. The method establishes entropy as a purely informational quantity that can be derived from raw simulation data alone, opening a new computational paradigm for evaluating free energies and phase equilibria across materials systems.%}
\end{abstract}

\maketitle

%\section{Introduction}

Entropy is the fundamental quantity linking information, disorder, and energy, yet its physical and informational interpretations have remained largely separate. 
Since the late nineteenth century, entropy has been defined as a measure of microscopic disorder ($S=k_B \ln \Omega$), first proposed by Boltzmann \cite{boltzmann} and later generalized by Gibbs \cite{Gibbs2011-qz} to nonuniform probability distributions ($S=-k_\mathrm{B}\sum_i p_i \ln p_i$).  A century later, Shannon introduced an analogous measure of information content \cite{shannon}, formalizing the minimal number of bits required to describe a system of symbols with given probabilities. At von Neumann’s suggestion \cite{SchementRuben1993}, he called it entropy ($H = -\sum_i p_i \log_2 p_i$), reflecting its mathematical identity with the Boltzmann–Gibbs form. Jaynes recognized their formal equivalence \cite{JaynesFirstPaper,JaynesSecondPaper}, showing that thermodynamic and information entropies differ only by a constant factor, $S = k_B \ln(2) H_{\mathrm{Shannon}}$. This deep connection suggests that the thermodynamic entropy of matter might, in principle, be obtained directly from its informational content.

Here we make this idea operational.
We demonstrate, for the first time for real condensed-matter systems, that Shannon’s information entropy and the thermodynamic entropy of Boltzmann and Gibbs are quantitatively equivalent.  Using density functional theory (DFT) molecular dynamics (MD) simulations, we compute the entropy of solids and liquids directly from the information content of their atomic configurations—without invoking any physical decomposition or empirical model.  
This establishes (1) a literal and operational equivalence between information and thermodynamic entropy for real condensed-matter systems, and (2) a fast, general, and parameter-free framework for computing entropies and free energies across diverse classes of materials.

While entropy connects the microscopic world of atomic configurations to the macroscopic laws of thermodynamics, its evaluation in real materials remains both computationally demanding and conceptually opaque.  Traditional approaches decompose the total entropy into vibrational, configurational, electronic, and magnetic parts through coarse-graining—each requiring distinct models, approximations, and sampling schemes \cite{Stot1,Stot2,Stot3,Stot4,Stot5,Stot6,Stot7,WidomGao2019}.  
Here we take a fundamentally different route: By compressing atomic configurations from DFT MD trajectories, we determine the minimal number of bits required to describe a microstate of atomic positions within a defined physical precision.  This bit count of information entropy maps directly to the thermodynamic entropy through the Shannon–Boltzmann relation, providing a parameter-free and model-independent route to entropy from simulation data alone.  Applied to metals, semiconductors, and refractory ceramics, in both solid and liquid phases, this compression-based framework reproduces benchmark entropies from SLUSCHI/mds \cite{SLUSCHI,Hong2025-ff} methods and demonstrates that thermodynamic entropy is information.

We implement this idea by treating the atomic configuration from a DFT MD trajectory as a message to be optimally compressed.  
%We demonstrate that entropy can be obtained directly from the minimum number of bits required to describe the system’s microstate.  
A microstate with correlated atomic motions can be “compressed” by removing redundancies, revealing its intrinsic informational content.  
Using a custom K-SVD \cite{KSVD} style compression scheme, we determine the minimum bit length required to encode a microstate and map that bit count to thermodynamic entropy via Eqs. \ref{eq:thermo_shannon_equiv} and \ref{eq:thermo_bits_equiv}.  
This approach, termed \textit{atomic entropy determination via first-principles} (ASDF), provides a direct, parameter-free route to entropy from first principles and information theory.

\begin{figure}
\includegraphics[width=0.49\textwidth]{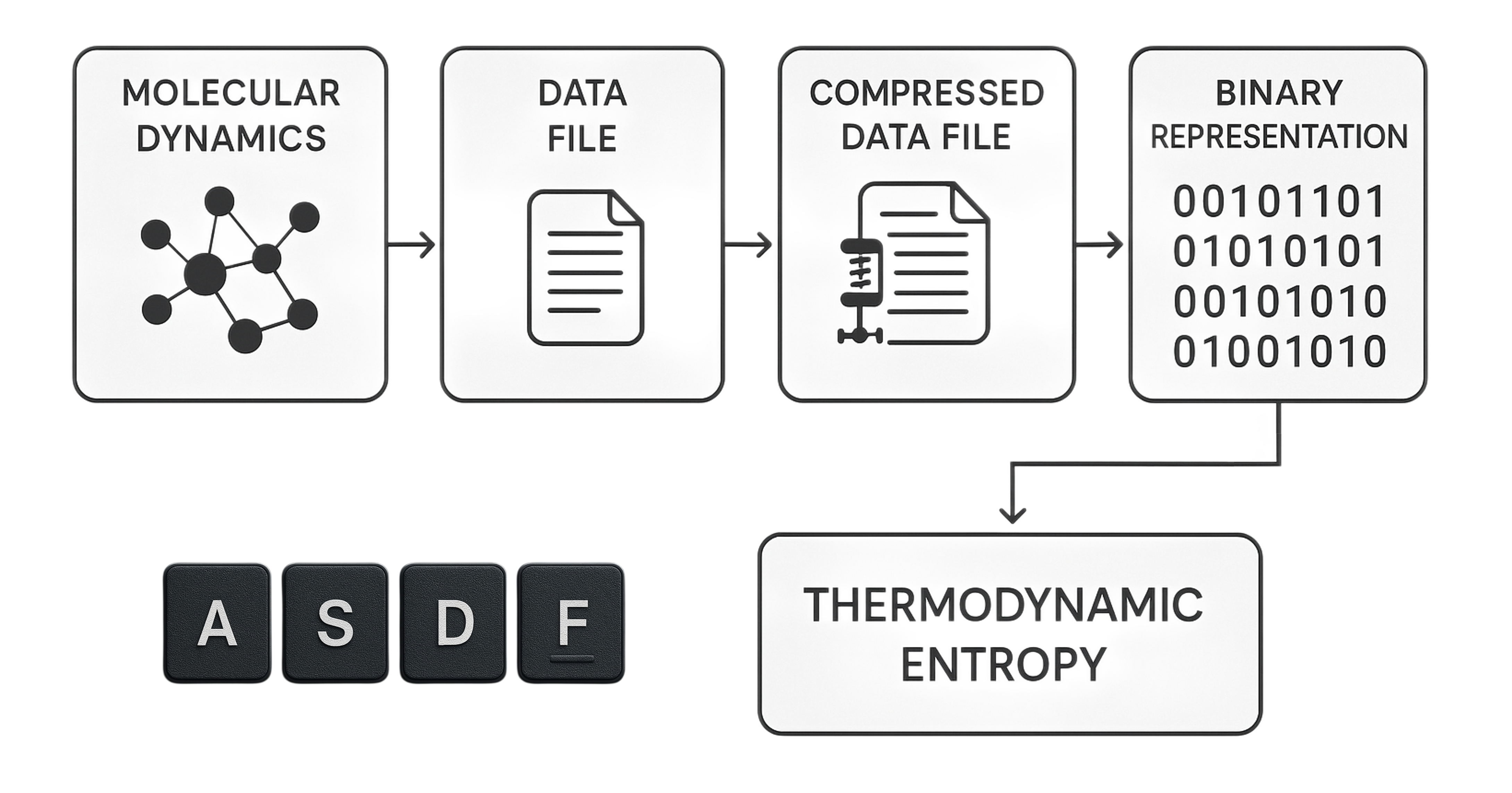}
\caption{\justifying \textbf{High level overview of asdf.} MD simulations are performed and the atomistic trajectory is reduced to the essential data describing each microstate. This data file, consisting of N atomic positions, is then compressed using a K-SVD–style sparse-dictionary algorithm, producing a compact representation that is converted into a bit string. The length of this bit string corresponds directly to the thermodynamic entropy of the microstate.}
\label{fig:placeholder}
\end{figure}

%\subsection*{Interpretation}

This procedure may be interpreted as a communication problem: a “sender’’ describes a microstate to a “receiver’’ who already knows the macroscopic conditions $(N,P,T)$. The language by which the sender and receiver communicate is analogous to the already agreed upon microstate to which all other microstates may be constructed with respect to.
The compressed bit string represents the minimal number of binary questions required to specify all atomic positions within the physical uncertainty $\varepsilon$ from quantum mechanics, analogous to transmission through a noisy channel limited by the de Broglie wavelength.  
By equating this irreducible information with the thermodynamic entropy via Eq. \ref{eq:thermo_per_mol}, ASDF provides a direct, parameter-free and phase-independent method to compute entropy from first principles.

%\section{Results}

The goal is to determine the minimal information content needed to reconstruct the mapping $\{\Delta\mathbf{r}_i\}$.  
We seek a sparse dictionary of basis vectors $B=\{\mathbf{b}_j\}_{j=1}^M$ and a discrete coefficient matrix $C\in\{-1,0,1\}^{M\times N}$ such that
% Discrete dictionary learning with per-row error control
\begin{equation}\label{eq:KSVD}
\min_{B,C}\big(M,\,H_\mathrm{coeff}(C)\big)
\quad \text{s.t.} \quad 
\|(\Delta R - BC)_{i}\|_2 \le \varepsilon \;\; \forall\, i,
\end{equation}
where $\Delta R=[\Delta\mathbf{r}_1,\ldots,\Delta\mathbf{r}_N]$ and $\varepsilon$ is the reconstruction error threshold.  
Each coefficient encodes an on/off or directional activation of a mode, forming a binary symbol stream whose entropy reflects the information required to describe the microstate.  

The value of $\varepsilon$ determines the resolution of physically meaningful information.  
Choosing $\varepsilon$ near or below the thermal de Broglie wavelength volume,
%\begin{equation}\label{eq:deBroglie}
%$\lambda_\mathrm{dB} = \frac{h}{\sqrt{3}\sqrt{2\pi m k_\mathrm{B}T}}$,
$\lambda_\mathrm{dB} = h / (\sqrt{3}\sqrt{2\pi m k_\mathrm{B}T})$,
%\end{equation}
ensures that quantum uncertainty limits are respected; further refinement below this scale captures negligible physical information, which is demonstrated in the results section. Decreasing the reconstruction threshold beyond the system’s intrinsic information resolution introduces artificial entropy growth, as the algorithm begins encoding numerical noise rather than new physical information. In this regime, the apparent entropy increases linearly,
%\begin{equation}\label{eq:S_artificial_increase}
$H(\varepsilon)=d\ln{1/\varepsilon}$,
%\end{equation}
with slope $d$ corresponding to the information dimensionality of the space \cite{information_dimension_original,information_dimension_ref}. In the results section, $d$ is shown to be phase-independent, thus preserving entropy difference $\Delta S$ (such as fusion entropy) with superfluous precision.

To evaluate whether the proposed equivalence holds universally across condensed-phase materials,
we test whether the average Shannon-entropy limit of microstate compressibility in equilibrium molecular-dynamics trajectories corresponds quantitatively to the thermodynamic entropy of each material.
We applied the information-theoretic framework to DFT MD data for a wide variety of materials, including metals, oxides, molten salts, and refractory ceramics. %aluminum, silicon, titanium, and tungsten. 
Each material was simulated in both solid and liquid phases near its experimental melting temperature, %---1000~K for Al, 1400~K for Si, 1800~K for Ti, and 3400~K for W---
using the SLUSCHI interface with VASP \cite{VASP1,VASP2,VASP3}. These systems span a wide range of bonding characteristics, atomic masses, and structural symmetries, providing a rigorous test of the method’s generality.

\begin{figure*}
    \includegraphics[width=0.9\linewidth]{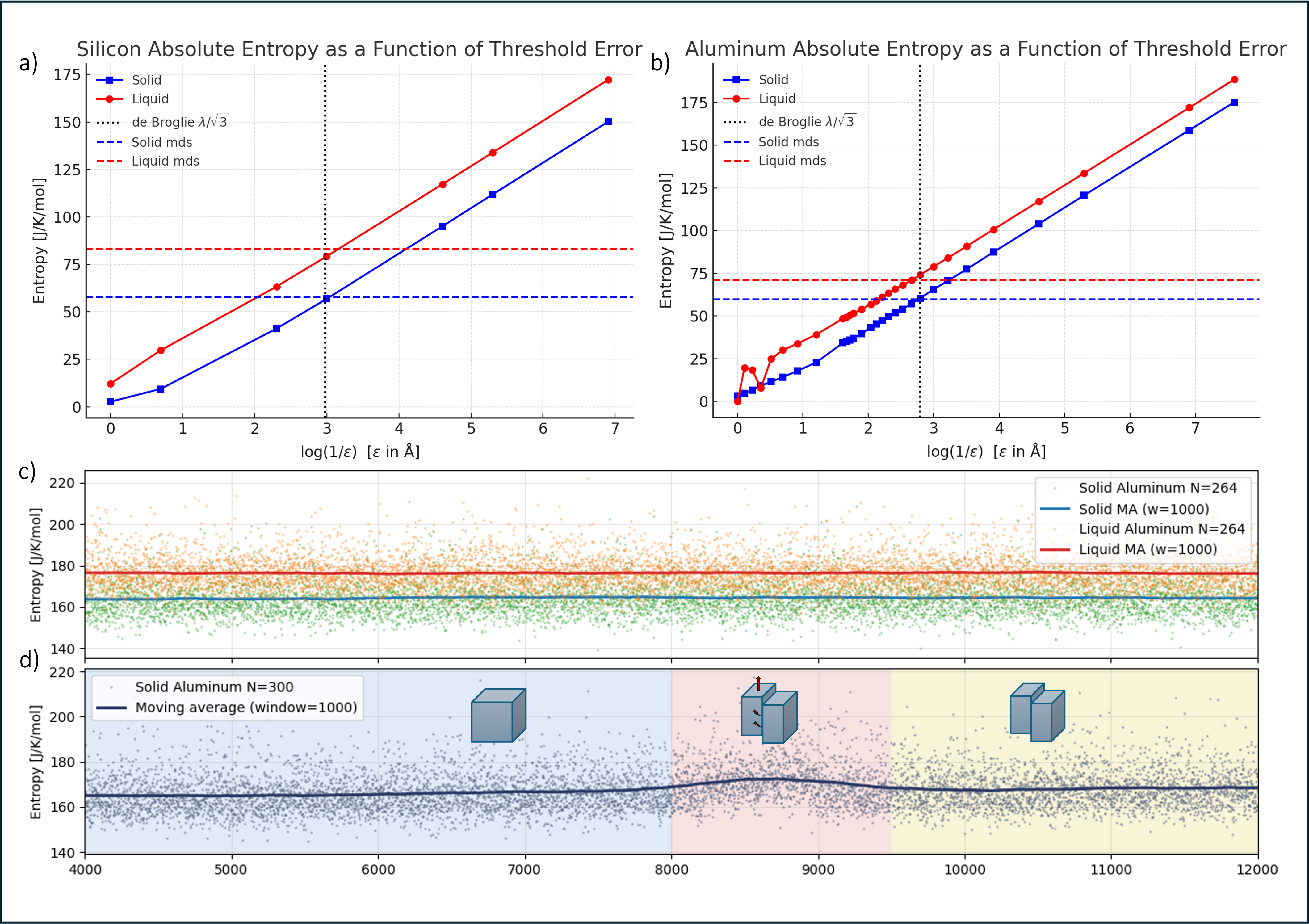}
    \caption{\justifying \textbf{Threshold-error behavior, stability, and trajectory sensitivity of the ASDF entropy method.} 
    (a) Absolute entropy of silicon as a function of $\log(1/\varepsilon)$. In information theory, artificially enlarging the threshold error $\varepsilon$ inflates the inferred information content proportionally to the system's ``information dimensionality.'' Since the thermal de~Broglie wavelength provides an approximate scale at which all physically relevant information is captured, the entropy exhibits a linear regime for $\varepsilon < \lambda_{\mathrm{dB}}$ with an effective dimensionality close to $3R$, nearly identical for solid and liquid silicon. Thus, the entropy value nearest the de~Broglie wavelength corresponds most closely to classical benchmarks. 
    (b) Same analysis for aluminum, exhibiting analogous behavior. 
    (c) Stability of the absolute entropy of solid and liquid aluminum over an extended MD trajectory. For the ASDF method to be well-defined, the entropy must be insensitive to trajectory length and to the specific segment sampled after an initial relaxation period; the figure confirms this stability. 
    (d) Sensitivity of the algorithm to rare or atypical configurations. During a section of one trajectory, a stacking-fault dislocation temporarily increases the measured entropy. Excluding such rare events yields a clean and consistent bulk entropy value. 
    \textit{Together, panels (a--d) demonstrate that the ASDF method is stable with respect to threshold error, trajectory length, and sampling window, while remaining sensitive enough to identify anomalous microstructural events.}}
    \label{fig:2}
\end{figure*}

%All entropies were evaluated from the Shannon expression applied to the coefficient matrix of the compression algorithm, corresponding to the minimal binary representation of the residual mapping between pairs of microstates and thus represents the theoretical upper bound on the information content of the system within the threshold error constraint. The resulting quantity, $H_\mathrm{coeff}$, was converted to thermodynamic entropy via equation \ref{eq:thermo_per_mol}, yielding the smallest achievable entropy consistent with the physical information of the trajectory.

All entropies were evaluated from the Shannon expression applied to the coefficient matrix of the compression algorithm. This matrix represents the minimal binary encoding of the residual atomic displacements between pairs of microstates. The resulting Shannon entropy, $H_\mathrm{coeff}$, quantifies the irreducible information content of the system at a given resolution and thus provides the smallest achievable entropy consistent with the physical information in the trajectory. $H_\mathrm{coeff}$ was converted to thermodynamic entropy using Eq. \ref{eq:thermo_per_mol}, establishing a direct mapping between the informational compressibility of atomic configurations and the material’s thermodynamic entropy.

%\subsection*{\texorpdfstring{$\Delta S$}{∆S} is invariant across computational parameters}

A central requirement of the ASDF method is that the reconstructed entropy be stable with respect to the coarse-graining threshold $\varepsilon$ and the entropy extracted from compressed microstates behaves as a physically meaningful quantity rather than a numerical artifact.
As shown in Figure~\ref{fig:2} (a) and (b) for Al and Si in both solid and liquid phases, the ASDF framework satisfies this condition decisively.
When the reconstruction threshold $\varepsilon$ is small (equivalently, when $\log(1/\varepsilon)$ is large), the extracted entropy increases linearly with $\log(1/\varepsilon)$ - a well-known manifestation of artificial information generated by oversampling a system at a resolution finer than the physically meaningful scale \cite{information_dimension_original}. This behavior can also be viewed as analogous to assigning an artificially larger atomic mass, since both de Broglie wavelength and phonon frequency scale as $1/\sqrt{m}$.
The slope of the linear regime corresponds to an effective information dimensionality close to $3R$, identical for both solid and liquid phases and consistent with the entropy of a classical harmonic oscillator.
Once $\varepsilon$ approaches $\lambda_{\mathrm{dB}}/\!\sqrt{3}$, this superficial inflation disappears, and the information entropy values agree closely with the thermodynamic entropy obtained from SLUSCHI/mds calculations.  
This agreement confirms that the de~Broglie wavelength marks the resolution at which nearly all physically resolvable information has been captured — no further detail increases the system’s entropy.% and below the de~Broglie wavelength the entropy converges linearly with respect to excessive precision.  
The same characteristic behavior is observed across all materials studied, demonstrating that the de Broglie wavelength defines a universal informational resolution for condensed-phase systems.

\begin{figure*}
    \includegraphics[width=1\linewidth]{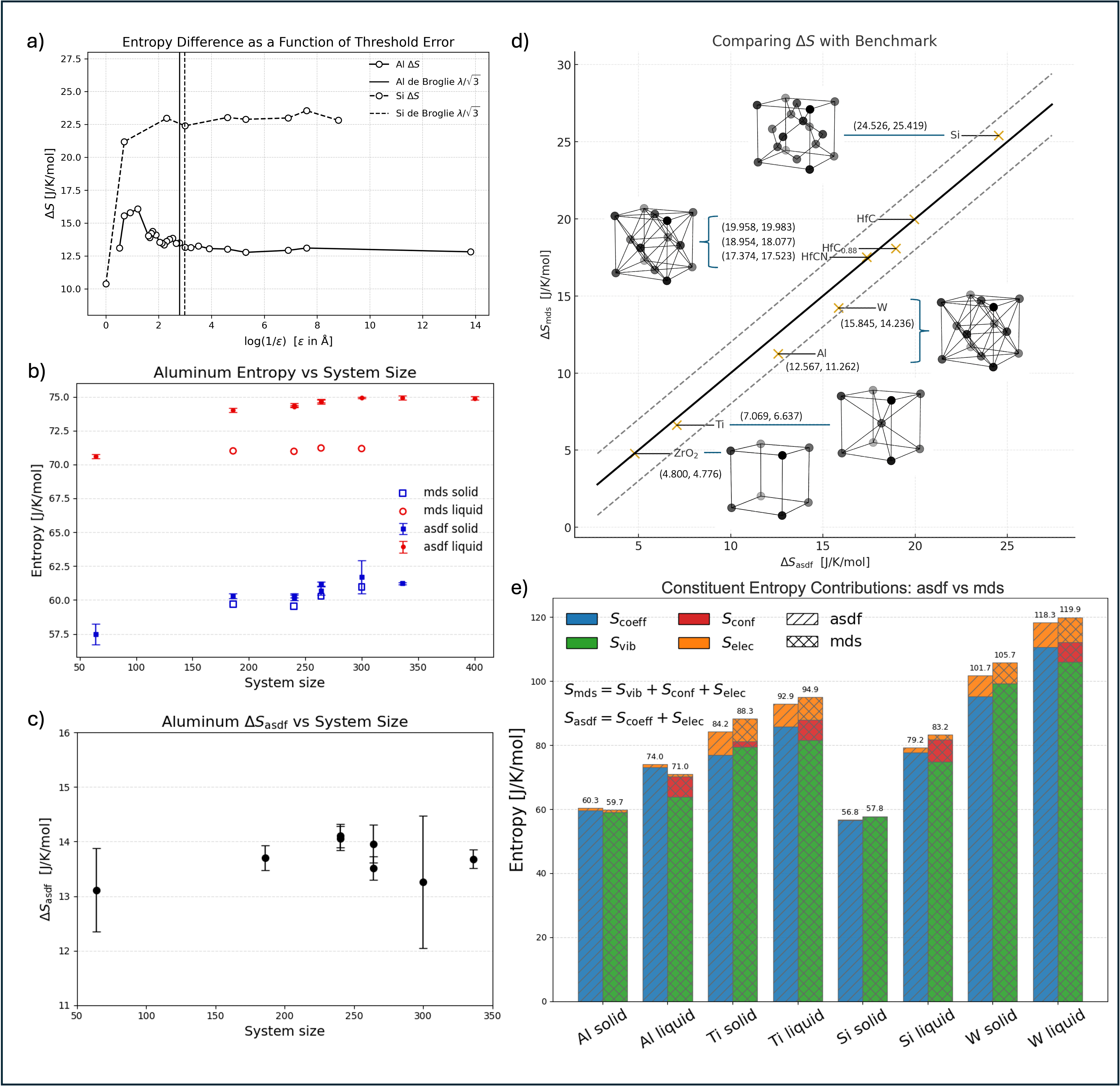}
    \caption{\justifying \textbf{System-size convergence, threshold-error independence, and agreement with benchmark entropy calculations.}
    (a) Stability of the entropy difference $\Delta S$ for silicon and aluminum as a function of $\log(1/\varepsilon)$. The plateau near the thermal de~Broglie wavelength confirms that $\Delta S$ becomes independent of the threshold error once all physically relevant information is resolved.
    (b) Absolute entropy of solid and liquid aluminum as a function of system size $N$, shown for both ASDF and the benchmark \texttt{mds} method. A minimum system size is required to capture bulk configurational complexity; beyond this, the entropy saturates and remains stable with increasing $N$. Both approaches exhibit a slight upward trend at large $N$, reflecting the increased configurational space sampled in larger systems.
    (c) Comparison of ASDF and \texttt{mds} entropies for Si, W, Al, and Ti using a tight threshold $\varepsilon = 10^{-4}$~\AA. This value minimizes threshold-induced error while remaining above the numerical noise floor of MD trajectories. All materials agree with \texttt{mds} to within $2~\mathrm{J\,K^{-1}\,mol^{-1}}$, demonstrating excellent precision.
    (d) Absolute solid and liquid entropies for the same four materials using $\varepsilon = \lambda_{\mathrm{dB}}$.
    ASDF combines coefficient entropy with the electronic entropy, whereas \texttt{mds} sums vibrational, electronic, and configurational contributions. 
    All materials show close agreement between the two methods.
    \textit{Together, panels (a--d) demonstrate independence from threshold error, convergence with system size, and excellent agreement of both $\Delta S$ and absolute $S$ with benchmark calculations.}
    }
    \label{fig:3}
\end{figure*}

Beyond threshold stability, the ASDF entropy must also remain constant over time for equilibrium systems while being sensitive to genuine structural fluctuations.
As illustrated in Figure~\ref{fig:2} (c) and (d), across extended molecular-dynamics trajectories, the entropy of both solid and liquid phases remains stable once the system has equilibrated, confirming that the compressed information content is an intrinsic property of the thermodynamic ensemble rather than a transient feature of a specific configuration.
Occasional deviations reveal the method’s diagnostic power: during one aluminum trajectory, the appearance of a stacking-fault dislocation produced a transient entropy spike before returning to the equilibrium value.
Such events demonstrate that ASDF distinguishes true microstructural changes from numerical noise, capturing the emergence of local disorder without affecting the long-term average.
Together, these behaviors establish that the ASDF entropy is both physically grounded and dynamically robust—stable for equilibrium systems, yet responsive to the rare fluctuations that carry thermodynamic significance. 

%Figure~\ref{fig:2}(c) demonstrates that the absolute entropy of solid and liquid aluminum remains stable over long MD trajectories once an initial relaxation period is discarded.  
%This stability is essential: the compressed information content must be independent of the precise segment of trajectory selected, provided the system is in equilibrium.  
%Panel~\ref{fig:2}(d) further illustrates an important diagnostic feature of the algorithm.  
%During a portion of one trajectory, a stacking-fault dislocation transiently increases the measured entropy.  
%The method therefore detects anomalous or non-equilibrium configurations, ensuring that bulk entropy can be extracted by sampling only the regions of the trajectory representing typical microstates of the phase.  

To verify that the information-derived entropy remains consistent across different system sizes and computational parameters, we examined its convergence behavior and relative stability between phases.
As the number of atoms $N$ increases, the information entropy obtained from ASDF approaches a well-defined bulk limit at approximately 200 atoms, reflecting the complete configurational complexity of the system, as shown in Figure~\ref{fig:3}(b). 
Beyond this point, both ASDF and benchmark SLUSCHI/mds calculations show only a slight upward drift, attributable to the longer wavelength phonons captured in extended supercells.
The persistence of this trend across all materials confirms that the ASDF information entropy scales extensively with system size and encodes the same underlying physics as conventional statistical-mechanical treatments.
Equally important, the entropy difference $\Delta S$ between solid and liquid phases is nearly invariant once the coarse-graining threshold satisfies $\varepsilon \lesssim \lambda_{\mathrm{dB}}$, as shown in Figure~\ref{fig:3}(a).
%Because the momentum contributions cancel between phases at identical temperature, only the configurational entropy governs $\Delta S$, which converges rapidly once all physically meaningful information is resolved.
This invariance indicates that ASDF captures relative entropies, such as fusion entropy $\Delta S$, with high precision and without parameter tuning, providing a consistent measure of entropy changes across phase boundaries and chemical systems.

%Figure~\ref{fig:3}(b) shows the absolute entropy of solid and liquid aluminum as a function of system size $N$ computed using both ASDF and the benchmark \texttt{mds} method.  
%A minimum system size is required to capture the complete configurational complexity of the bulk phase; beyond this value, the entropy saturates and remains effectively constant.  
%Both ASDF and \texttt{mds} exhibit a similar slight upward drift with increasing $N$, reflecting the increased diversity of local configurations in larger supercells.  
%This agreement reinforces that ASDF captures the same physics encoded in conventional configurational and vibrational sampling techniques.

%Figure~\ref{fig:3}(a) shows that the entropy difference $\Delta S$ for aluminum and silicon is essentially independent of threshold error once $\varepsilon \lesssim \lambda_{\mathrm{dB}}$.  
%This behavior is expected: since the momentum contribution is identical for both phases at the same temperature, only the configurational entropy differs, and this difference converges rapidly once all physically meaningful information is resolved.  
%Thus, ASDF provides a robust method for extracting relative entropies without requiring precise tuning of $\varepsilon$.

%\subsection*{Agreement with benchmark}
The consistency of the ASDF framework across diverse materials further demonstrates its generality.
As shown in Figure~\ref{fig:3}(c) When applied to metals, semiconductors, oxides, and refractory ceramics, the information entropies derived from ASDF agree quantitatively with benchmark SLUSCHI/mds results within 2 J$\cdot$K$^{-1}$$\cdot$mol$^{-1}$—less than a 2\% deviation even under stringent precision thresholds using $\varepsilon = 10^{-4}$~\AA \,\,to achieve precision.
When using the de~Broglie wavelength, the absolute solid and liquid entropies for the materials are recovered, as illustrated in Figure~\ref{fig:3}(d).  
In ASDF, the total entropy arises naturally from the informational content of the compressed microstates, combining configurational and vibrational contributions without explicit physical decomposition.
Despite this fundamentally different formulation, the method reproduces the same absolute and relative entropies as conventional approaches that sum vibrational, configurational, and electronic terms.
This numerical equivalence confirms that thermodynamic entropy can be obtained directly from the informational compressibility of atomic configurations, establishing information theory as a complete and physically grounded route to the thermodynamics of condensed matter.

%%Figure~\ref{fig:3}(c) compares ASDF and \texttt{mds} entropies for Si, W, Al, and Ti using $\varepsilon = 10^{-4}$~\AA.  
%This tight threshold reduces statistical variation while remaining above the numerical noise floor of the MD trajectories.  
%All materials agree with the benchmark values to within $2~\mathrm{J\,K^{-1}\,mol^{-1}}$, highlighting the precision of the method even under stringent resolution.  
%ASDF combines the coefficient entropy with the electronic entropy, whereas \texttt{mds} combines vibrational, electronic, and configurational terms.  
%Agreement across all materials confirms the validity of using a purely information-theoretic representation to compute thermodynamic quantities traditionally obtained through physically distinct contributions.

Taken together, these results demonstrate that thermodynamic entropy can be derived directly from the minimal number of bits required to describe a microstate under fixed macroscopic conditions.
No enumeration of microstates, partitioning into vibrational or configurational components, or physical modeling assumptions are needed beyond the molecular-dynamics trajectory itself.
By compressing the mapping between microstates within an equilibrium ensemble and invoking the Shannon–Boltzmann relation, ASDF is an accurate and fully general method that yields both absolute and relative entropies with first-principles accuracy.
This demonstrates, in a literal and operational sense, that thermodynamic entropy \emph{is} Shannon entropy: the measure of the irreducible information needed to specify a microstate.  
This reinterpretation reframes entropy not as a derived thermodynamic quantity but as a directly measurable property of information storage within atomic configurations.

%The quantitative agreement across materials and the invariance of $\Delta S$ with respect to system size, threshold precision, and material class reveal a deeper principle: the informational compressibility of matter mirrors its thermodynamic disorder.
%This equivalence provides a unified foundation linking information theory and statistical mechanics, while also offering a practical and computationally efficient route for evaluating entropies and free energies in complex materials.
%Because the ASDF method relies solely on raw trajectory data, any algorithm capable of reconstructing a microstate defines an upper bound that converges monotonically toward the true thermodynamic entropy—since redundant information can only increase the measured bit count.
%In doing so, this work bridges two historically distinct concepts—information and energy—within a single quantitative framework.
Together, these results establish that the thermodynamic entropy of a condensed system can be computed directly from the average compressibility of its microstates, without recourse to traditional physical partitioning or empirical parameters. The observed invariance of $\Delta S$ with respect to threshold error, system size, and material type demonstrates that the relationship is invariant of extensive parameters. 
The quantitative agreement between the information-theoretic and thermodynamic entropies observed here suggests that the two measures are not merely analogous in form but arise from the same underlying constraint on distinguishability within phase space--linking the compressibility of information to the physical disorder of matter. Practically, this equivalence provides a purely informational route to entropy: because no physical modeling is required, any algorithm capable of reconstructing a microstate defines an upper bound that monotonically converges toward the true thermodynamic entropy, since redundant information can only increase the measured bit content. This convergence of statistical mechanics and information theory provides a unified and computationally efficient foundation for calculating entropy from first principles.

%Taken together, Figures~\ref{fig:2} and~\ref{fig:3} show that thermodynamic entropy can be obtained directly from the minimal number of bits required to describe a microstate at fixed macroscopic conditions.  
%No enumeration of microstates is required.  
%No partitioning into vibrational, configurational, or electronic components is required.  
%No physical modeling assumptions are required beyond the MD trajectory itself.  
%By compressing the mapping from one microstate to another within the same equilibrium ensemble---and by taking seriously the identity in %Eq. \ref{eq:thermo_bits_equiv}
%we obtain an accurate and fully general method for computing both absolute and relative entropies.  
%The results highlight the conceptual value of treating thermodynamic entropy as an information-theoretic quantity and suggest that compression algorithms, rather than traditional physical decompositions, may provide an alternative route for understanding and computing entropy in complex materials.

\bibliography{citations}% common bib file
%% if required, the content of .bbl file can be included here once bbl is generated
%%\input sn-article.bbl

%\backmatter

%\bmhead{Supplementary information}

%\bmhead{Acknowledgments}

%%===========================================================================================%%
%% If you are submitting to one of the Nature Portfolio journals, using the eJP submission   %%
%% system, please include the references within the manuscript file itself. You may do this  %%
%% by copying the reference list from your .bbl file, paste it into the main manuscript .tex %%
%% file, and delete the associated \verb+\bibliography+ commands.                            %%
%%===========================================================================================%%
\section{\ Methods}

\subsection*{Data generation and ensemble definition}

Ab-initio molecular-dynamics simulations of NPT ensembles were performed using the SLUSCHI \cite{SLUSCHI} package interfaced with VASP \cite{VASP1,VASP2,VASP3}, employing the Perdew–Burke–Ernzerhof \cite{PBE} exchange–correlation functional with the projector augmented-wave (PAW) \cite{PAW} method. 
Each ionic step in the trajectory corresponds to an accessible microstate constrained by the macroscopic quantities $(N, P, T)$, each of which is compressed in parallel \cite{super_computer_parallel,sol}.  
A snapshot at time $t_i$ therefore represents a point in the $6N$-dimensional phase space $(\mathbf{q}_N,\mathbf{p}_N)$, where $\mathbf{q}_N$ and $\mathbf{p}_N$ are the sets of atomic positions and momenta, respectively.  
If the microstate could be encoded into its smallest possible binary representation, the number of bits required for any shortest lossless description of the microstate would correspond to its thermodynamic entropy through Eq. \ref{eq:thermo_bits_equiv}.

\subsection*{Relative entropy and reference microstate}

Entropy is inherently a relative quantity; only differences $\Delta S$ enter measurable thermodynamic relations such as
\begin{equation*}\label{eq:G_and_Tm}
G = H - TS, \qquad T_\mathrm{m} = \frac{\Delta H}{\Delta S}.
\end{equation*}
It is customary to define $S=0$ at absolute zero, consistent with Clausius’ formulation, but in practice any well-defined reference can serve.  
Here, we replace the 0 K lattice by a reference microstate from the same equilibrated trajectory. This ensures identical $S=0$ reference, allowing a disregard for quantum or residual information adding unnecessary complexity and ambiguity, while preserving $\Delta S$.
For each target microstate, a reference is selected at random from snapshots sufficiently separated in time to allow thermalization, eliminating non-ergodic temporal correlations.  
The positional mapping between the two defines a set of $N$ three-dimensional residual vectors,
\begin{equation*}\label{eq:residual_map}
\Delta \mathbf{r}_i = \mathbf{r}_i^{(\mathrm{target})} - \mathbf{r}_{\epsilon(i)}^{(\mathrm{ref})},
\end{equation*}
where $\epsilon(i)$ indexes the nearest atom in the reference configuration.  
Averaging over many such pairs yields the mean information required to reconstruct one microstate from another under identical macroscopic constraints.

\subsection*{Neglect of momentum-space contributions}

The total entropy separates into kinetic and configurational components,
\begin{equation*}\label{eq:S_Skin_Sconf}
S = S_\mathrm{kin} + S_\mathrm{conf}.
\end{equation*}
For a canonical ensemble, the kinetic term is
\[
S_\mathrm{kin} = \frac{3}{2}Nk_\mathrm{B}\!\left[1 + \ln\!\left(\frac{2\pi m k_\mathrm{B}T}{h^2}\right)\right],
\]
which depends only on $T$ and $m$.  
Because the solid and liquid phases share identical Maxwell–Boltzmann momentum distributions at fixed $T$, $S_\mathrm{kin}$ cancels in $\Delta S$.  
Hence, only the configurational subspace $\mathbf{q}_N$ is considered.

%\subsection*{Compression of configurational information}

\subsection*{K-SVD–based dictionary learning}
The basis $\mathbf{B}$ and coefficients $\mathbf{C}$ are obtained through an iterative sparse-dictionary learning procedure analogous to the K–SVD algorithm \cite{KSVD}, which in general seeks to minimize the reconstruction error
in Eq.\ref{eq:KSVD}.
K–SVD alternates between solving for the sparse coefficients at fixed basis vectors and updating each basis vector via singular-value decomposition to optimally reduce the Frobenius-norm residual.  
In the present work, we follow the same fundamental objective but impose an additional physical constraint: the coefficients are restricted to the discrete ternary set $\{-1,0,1\}$.  
This restriction both reflects the directional character of atomic displacements and enables direct evaluation of the Shannon entropy from the resulting bit-string representation.

At each iteration, the covariance matrix of residual vectors,
\[
\boldsymbol{\Sigma} = \frac{1}{N}\,\Delta\mathbf{R}\,\Delta\mathbf{R}^{\mathsf{T}},
\]
is diagonalized to obtain the dominant eigenvector, defining the next basis vector $\mathbf{b}_1$.  
Its optimal scaling is determined by identifying the knee point in the cumulative distribution of residual norms, maximizing the reduction in mean-square error.  
Residuals whose magnitudes decrease under this projection are updated, and those below the threshold $\varepsilon$ are removed.  
The procedure is repeated on the remaining residuals until all are within the prescribed tolerance.  
Finally, a relaxation step re-optimizes all coefficients simultaneously over the complete basis set to remove redundancy and yield a compact dictionary suited for subsequent entropy evaluation.

\subsection*{Entropy decomposition and scaling}

The total information content of the compressed representation can be written as
\[
H_\mathrm{total}
   = H_\mathrm{alg} + H_\mathrm{basis} + H_\mathrm{coeff},
\]
where \(H_\mathrm{alg}\) is the information required to specify the encoding instructions (quite literally Eq. \ref{eq:KSVD} itself),
\(H_\mathrm{basis}\) the information needed to store the basis vectors, and
\(H_\mathrm{coeff}\) the Shannon entropy of the coefficient sequence.
In this view, \(H_\mathrm{total}\) approximates the Kolmogorov complexity \cite{Kolmogorov1998-ii}
of the microstate \(x\)—that is, the length of the shortest effective description
from which the microstate can be reconstructed.

More formally, Kolmogorov complexity decomposes into contributions from the
model (the algorithm and basis) and the data conditioned on that model
(the coefficients),
which mirrors the minimum-description-length (MDL) decomposition in algorithmic
information theory \cite{MDL}.
Accordingly, the Shannon term \(H_\mathrm{coeff}\) estimates the conditional Kolmogorov
complexity of the microstate given the chosen compression model, while
\(H_\mathrm{alg} + H_\mathrm{basis}\) quantify the complexity of the model itself.
The sum therefore provides a finite, operational approximation to the total
algorithmic information contained in the microstate.

Liu and Simine recently demonstrated that molecular binding affinities can be approximated by relating the Kolmogorov complexity of molecular configurations to their thermodynamic entropy through a lossless compression framework \cite{KolmogorovChemistry}.  
Their results highlight the broader potential of information–theoretic measures in modeling physical systems, and motivate the more general approach developed in this work. Several prior studies have employed Kolmogorov complexity in physical modeling \cite{KolmogorovInPhysics1,KolmogorovPhysics2}, suggesting a deeper underlying connection between information measures and thermodynamic behavior.

In our K-SVD style approach, the algorithmic complexity is the information needed to encode the mapping of Eq. \ref{eq:KSVD}, and does not scale with N.
Therefore, only $H_\mathrm{coeff}$ scales linearly with system size; the others remain constant or sublinear.  
In the molar limit,
\begin{equation}\label{eq:thermo_per_mol}
S\,[J K^{-1}\mathrm{mol}^{-1}] \approx \frac{N_A}{N} k_\mathrm{B}\ln(2)\,H_\mathrm{coeff}
\end{equation}
where $N_A$ is the Avogadro constant, and N is simulation system size.
$H_\mathrm{coeff}$ can be computed directly from the empirical distribution of $\{-1,0,1\}$ entries,
\[
H_\mathrm{coeff} = -\!\!\!\!\!\!\!\sum_{x\in\{-1,0,1\}}\!\!\!\!\!\!p_x\log_2 p_x,
\]
or equivalently from the bit length of an optimally encoded binary stream (e.g., Huffman coding).

\section*{Declarations}
This research was supported by US Department of Defense Army Research Office Award number W911NF-23-2-0145, with use of Research Computing at Arizona State University. 

All data generated or analyzed during this study will be made publicly available upon publication of the manuscript.  
Processed datasets and numerical outputs underlying the figures will be deposited in an open-access repository at the time of publication.

The ASDF codebase used in this study, including all scripts required to reproduce the results, will be released in a public repository concurrent with publication.  
Instructions for reproducing entropy calculations and compression workflows will be provided alongside the code.

The authors declare no competing interests. \\ \vspace{19cm}

\pagebreak

\onecolumngrid
\begin{center}
\large{Supporting Information for\\
\textit{Thermodynamic Entropy as Information\\
\small A compression-based demonstration of the Shannon–Boltzmann equivalence in condensed matter}}
\end{center}

\twocolumngrid

\section{Appendix 1: History of Entropy}\label{secA1}

Throughout the history of physics, entropy has evolved as both a physical quantity and a conceptual idea. Rudolf Clausius first introduced the term in the mid-nineteenth century as a measure of the portion of a system’s energy unavailable to perform useful work, expressed in differential form as \cite{Clausius1865-gi}
\begin{equation*}\label{eq:claudius_entropy}
dS = \frac{\delta Q_{\mathrm{rev}}}{T},
\end{equation*}
where $\delta Q_{\mathrm{rev}}$ is the infinitesimal amount of heat exchanged reversibly at temperature $T$.  
Ludwig Boltzmann later provided a microscopic foundation, showing that entropy scales with the number of accessible microstates $\Omega$ available to a system \cite{boltzmann},
\begin{equation}\label{eq:boltzmann_entropy}
S = k_\mathrm{B}\ln\Omega,
\end{equation}
where $k_\mathrm{B}$ is Boltzmann’s constant, introduced to connect Clausius’ macroscopic units to the microscopic scale.  
Josiah Willard Gibbs generalized Boltzmann’s expression to systems described by a non-uniform probability distribution $\{p_i\}$ over microstates $i$ \cite{Gibbs2011-qz},
\begin{equation}\label{eq:gibbs_entropy}
S = -k_\mathrm{B}\sum_i p_i \ln p_i,
\end{equation}
establishing entropy as a measure of uncertainty or disorder within a statistical ensemble.

Nearly a century later, Claude Shannon introduced an analogous quantity while developing the mathematical theory of communication.  
For a discrete set of symbols $\{x_i\}$ occurring with probabilities $\{p_i\}$, he defined the mean information content per symbol as \cite{shannon}
\begin{equation*}\label{eq:shannon_entropy}
H = -\sum_i p_i \log_2 p_i,
\end{equation*}
which represents the minimum average number of bits required to encode a message after all redundancies have been removed.  
The structural similarity between Eqs. \ref{eq:boltzmann_entropy} and \ref{eq:gibbs_entropy} was immediately recognized.  
John von Neumann humorously advised Shannon to name his measure “entropy,” observing that “no one really knows what entropy is, so in a debate you will always have the advantage.” \cite{SchementRuben1993}
Von Neumann himself had already defined the quantum analogue of Gibbs’ entropy, the von Neumann entropy \cite{Von_Neumann1996-tm},
\begin{equation*}\label{eq:vonNeumann_entropy}
S_\mathrm{vN} = -k_\mathrm{B}\,\mathrm{Tr}(\hat{\rho}\ln\hat{\rho}),
\end{equation*}
where $\hat{\rho}$ is the density operator.  

This shared terminology blurred the line between thermodynamic and information entropy.  
The ensuing debate on their equivalence motivated new thought experiments and theoretical advances.  
Szilard’s one-molecule engine \cite{Szilard1929-pq} illustrated that information could be converted into mechanical work, and Landauer’s principle \cite{LaundauersPrinciple} quantified this link, stating that the erasure of one bit of information dissipates a minimum heat of
\begin{equation}\label{eq:launders_principle}
E_\mathrm{bit} = k_\mathrm{B}T\ln 2.
\end{equation}
This established a physical cost for information processing and grounded the Shannon measure in thermodynamic terms.  
Experimental realizations of Szilard-type engines \cite{LaundersExperimental} have since verified Eq. \ref{eq:launders_principle}, reinforcing the connection between information and energy.

E. T. Jaynes later demonstrated that Shannon and thermodynamic entropies are equivalent up to a unit conversion \cite{JaynesFirstPaper,JaynesSecondPaper},
\begin{equation}\label{eq:thermo_shannon_equiv}
S = k_\mathrm{B}\ln(2)\,H_\mathrm{Shannon},
\end{equation}
providing a quantitative bridge between bit-based and energy-based measures of uncertainty.  
Shannon argues $H$ represents the minimum achievable mean code length per symbol under any lossless encoding \cite{shannon}.  Equation \ref{eq:thermo_shannon_equiv} then implies that the thermodynamic entropy of a system is approximately equal to the bit length of its optimally compressed microscopic description,
\begin{equation}\label{eq:thermo_bits_equiv}
S = k_\mathrm{B}\ln(2)\,N_\mathrm{bits}.
\end{equation}

If one interprets the equivalence defined in equation \ref{eq:thermo_bits_equiv} literally, then it seems reasonable that one can compute the thermodynamic entropy directly from data compression.  
Traditional first-principles approaches, such as those used in DFT MD simulations, rely on partitioning the total entropy into independent physical components,
\begin{equation*}\label{eq:S_partition}
S = S_\mathrm{vib} + S_\mathrm{conf} + S_\mathrm{elec} + S_\mathrm{mag},
\end{equation*}
where each term describes vibrational, configurational, electronic, and magnetic contributions, respectively.

Numerous methods have been developed to evaluate these contributions independently, each tailored to the physical mechanisms governing the corresponding degrees of freedom.
Configurational entropy is commonly obtained through statistical enumeration of atomic arrangements \cite{Stot2,Stot3,Stot4,Stot5,Stot6,Stot7}. 
Vibrational entropy is typically computed from phonon density-of-states analyses or quasiharmonic approximations based on first-principles force constants \cite{LinBlancoGoddard2003,Frenkel2001}. 
Electronic entropy can be evaluated directly from the distribution of electronic occupations near the Fermi level within DFT or related frameworks. 
Collectively, these approaches have proven effective for specific classes of materials, yet they require distinct physical models, extensive sampling, and assumptions particular to each entropy term.

By contrast, if the equivalence in Eq. \ref{eq:thermo_bits_equiv} is taken at face value, entropy can be obtained directly from the minimum number of bits required to describe the system’s microstate.  
A microstate with correlated atomic motions can be “compressed” by removing redundancies, revealing its intrinsic informational content.  
In this work, we implement this idea by treating the atomic configuration from a DFT MD trajectory as a message to be optimally compressed.  
Using a custom K-SVD style compression scheme, we determine the minimum bit length required to encode a microstate and map that bit count to thermodynamic entropy via Eqs. \ref{eq:thermo_shannon_equiv} and \ref{eq:thermo_bits_equiv}.  
This approach, termed \textit{atomic entropy determination via first-principles} (ASDF), provides a direct, parameter-free route to entropy from first principles and information theory.

\section{Appendix 2: Separation of Kinetic and Configurational Contributions}\label{secA2}

Starting from the full \(N\)-particle canonical partition function
\[
\begin{aligned}
Z(&N,V,T)\\
&=\frac{1}{N!\,h^{3N}}
\int\!\mathrm{d}^{3N}p\;\mathrm{d}^{3N}r\;
\exp\Bigl[-\beta\bigl(K(p^N)+U(r^N)\bigr)\Bigr],
\end{aligned}
\]
momentum and configurational integrals may be factored:
\[
\begin{aligned}
Z(&N,V,T)\\
&=\frac{1}{N!\,h^{3N}}
\Bigl[\!\!\underbrace{\int_{\mathbb R^3} e^{-\beta\,p^2/(2m)}\,\mathrm{d}^3p}_{(2\pi m kT)^{3/2}}\Bigr]^{\!N}
\underbrace{\int_V e^{-\beta\,U(r^N)}\,\mathrm{d}^{3N}r}_{Q_N(V,T)}\\[6pt]
&=\frac{1}{N!}\,\frac{(2\pi m kT)^{3N/2}}{h^{3N}}\;Q_N(V,T)\\\\
\;&=\;\frac{1}{N!\,\lambda^{3N}}\,Q_N(V,T),
\end{aligned}
\]
where
\[
\lambda \;=\;\frac{h}{\sqrt{2\pi m kT}}
\quad\]
and
\[
Q_N(V,T)
=\int_V e^{-\beta\,U(r_1,\dots,r_N)}\,\mathrm{d}^3r_1\cdots\mathrm{d}^3r_N.
\]
The Gaussian momentum integral is the normalization constant of the Maxwell-Boltzmann distribution:
\[
f(\mathbf p)
\;=\;\Bigl(\tfrac{\beta}{2\pi m}\Bigr)^{3/2}
\exp\!\Bigl[-\beta\tfrac{p^2}{2m}\Bigr],
\]
\[
f(v)
=4\pi\Bigl(\tfrac{m}{2\pi kT}\Bigr)^{3/2}v^2\exp\!\Bigl[-\tfrac{m v^2}{2kT}\Bigr].
\]
Thus all of the phase-space information contained in the \(\mathbf p\)-coordinates is a function only of \(m\) and \(T\).

\vspace{0.5ex}
The Helmholtz free energy and entropy split accordingly:
\[
F=-kT\ln Z = F_{\rm kin}(T,m) + F_{\rm conf}(V,T), 
\]
\[
S=-\Bigl(\frac{\partial F}{\partial T}\Bigr)_{V,N}
=S_{\rm kin}(T,m) + S_{\rm conf}(V,T).
\]
Because \(F_{\rm kin}\) (and hence \(S_{\rm kin}\)) depends only on \(T\) and the particle mass, two phases of the same substance at the same \(T\) share identical momentum‐space entropy.  Therefore
\[
\Delta S \;=\;\Delta S_{\rm conf},
\]
and momentum contributions cancel in any entropy difference near the melting point.

\vspace{0.5ex}
The configurational integral \(Q_N\) must then be considered, and following Widom’s insertion-particle derivation \cite{WidomGao2019}:
\[
Q_N \;=\; Q_{N-1}\;V\;\bigl\langle e^{-\beta\Psi}\bigr\rangle.
\]
Instead of computing each microscopic interaction potential \(\Psi\) exactly, one can encode the ensemble into a sparse, lossy‐compressed basis whose dimension grows in proportion to the thermodynamic entropy.  Thus the size of the information content provides a direct measure of \(S_{\rm conf}\), as the paper aims to demonstrate.
 Projecting the full \(6N\)-dimensional phase-space trajectory onto the \(3N\)-dimensional configurational subspace preserves all entropy differences between phases at equal \(T\), justifying the neglect of momentum contributions.

\section{Appendix 3: Statistical Validity of Compression‐Based Entropy Estimates}

Let a single microstate snapshot \(n\) of our MD trajectory be described, under a maximum reconstruction error \(\epsilon\), by its optimal lossy‐compression size
\[
S_n(\epsilon)\,,
\]
and let our practical compressor produce measured sizes
\(\;S_n^p(\epsilon)\ge S_n(\epsilon)\).  By hypothesis
\[
\bigl\langle S_n(\epsilon)\bigr\rangle \;=\; S_{\rm therm}(\epsilon),
\]
and for two phases at the same \(T\) and \(\epsilon\), the entropy difference \(\Delta S_{\rm therm}\) is independent of \(\epsilon\).

The measured and physical entropy values can be modeled as two random variables:
\[
\begin{aligned}
S_n^p &\sim \Gamma\bigl(\alpha,\theta\bigr),\\
S_n   &\sim \mathcal{N}\!\bigl(S_{\rm therm},\,\sigma^2\bigr)\,,
\end{aligned}
\]
where \(\Gamma(\alpha,\theta)\) is the gamma distribution (shape \(\alpha\), scale \(\theta\)) and \(\mathcal N\) is a normal distribution peaked at the true thermodynamic entropy \(S_{\rm therm}\).

Define the deviation from the thermodynamic mean:
\[
\Delta_n \;=\; S_n^p \;-\; S_{\rm therm}\,.
\]
Its probability density is the convolution
\[
f_{\Delta}(\delta)
\;=\;
\int_{-\infty}^{\infty}
f_{\Gamma}(x)\;
f_{\mathcal N}(\delta - x)\,
dx.
\]
Hence the probability that a single compressed snapshot undershoots the theoretical limit is
\[
P\bigl(S_n^p < S_{\rm therm}\bigr)
=\;P(\Delta_n < 0)
=\;\int_{-\infty}^{0} f_{\Delta}(\delta)\,d\delta.
\]
For \(M\) independent snapshots, the probability that the minimum measured compression size
\(\min\{S_1^p,\dots,S_M^p\}\) falls below \(S_{\rm therm}\) is
\[
1 - \bigl[1 - P(S_n^p < S_{\rm therm})\bigr]^{M}.
\]

In the limit of infinitely dense phase-space sampling-when the true snapshot entropy distribution collapses to a Dirac delta (\(\sigma\to0\))-This gives
\(\mathcal N\to\delta(\cdot)\) and thus \(f_{\Delta}\to f_{\Gamma}\) with support \(\ge0\).  Consequently
\[
P\bigl(S_n^p < S_{\rm therm}\bigr)\;\longrightarrow\;0,
\]
and the minimum measured value approaches \(S_{\rm therm}\) from above.  

Under these statistical models, taking the lower bound of the domain of the gamma-distribution fitted to many lossy‐compression-based entropy estimates yields a statistically rigorous bound on the true thermodynamic entropy. 

\subsection{Coordinate System Considerations}

When compressing the residuals \(\{\Delta\mathbf{r}_i\}\), the goal is to exploit any underlying correlations so that a minimal dictionary suffices. Two key metrics guide the choice of coordinate frame:

\[
\mathbf{C}
= \frac{1}{2N}\sum_{i=1}^{2N}\Delta\mathbf{r}_i\,\Delta\mathbf{r}_i^T,
\]
\[
I (U;V)
= \iint p(u,v)\,\log\frac{p(u,v)}{p(u)\,p(v)}\,du\,dv,
\]

where \(\lambda_1\) (the largest eigenvalue of \(\mathbf{C}\)) measures how much variance a single PCA mode captures, and \(I (U;V)\) quantifies coupling between coordinates \(U\) and \(V\). Therefore, we choose the frame \(\mathcal{F}\) that simultaneously

\[
\text{maximizes }\lambda_1^{(\mathcal{F})}
\quad\text{and minimizes }\sum_{U\neq V} I _\mathcal{F}(U;V).
\]

In Cartesian coordinates, collective phonon modes often manifest as strong correlations between \(x\) and \(y\) (and similarly \(y\)-\(z\), \(z\)-\(x\)). For example, if half the atoms vibrate predominantly along an \(x\!-\!y\) plane phonon, the marginal distributions of \(x\) and \(y\) will show pronounced peaks in the same region, and the joint distribution will reveal high mutual information \(I (x;y)\). By encoding that phonon basis vector once, we avoid re-encoding its contribution for each atom.

Conversely, in spherical coordinates \((r,\theta,\phi)\), the radial coordinate \(r\) often carries most of the displacement information, while the angular distributions of \(\theta\) and \(\phi\) are nearly uniform. Their bar graphs are flat, and \(I (r;\theta)\approx I (r;\phi)\approx I (\theta;\phi)\approx0\). Overlaying the bar graphs for \(x,y,z\) versus those for \(r,\theta,\phi\) reveals that Cartesian axes share information, whereas spherical angles are essentially random fluctuations around a constant mean.

\end{document}